\documentclass[letter,twocolumn]{jpsj2}

\usepackage{graphicx}

\def\ve#1{\mbox{\boldmath $#1$}}

\def\Tc{T_\mathrm{c}}
\def\Re{\mathrm{Re}}

\title{%
Perturbation Theory for a Repulsive Hubbard Model \\
in Quasi-One-Dimensional Superconductors
}

\author{%
Sotaro \textsc{Sasaki}
Hiroaki \textsc{Ikeda}
and Kosaku \textsc{Yamada}
}

\inst{
Department of Physics, Kyoto University, Sakyo-ku, Kyoto 606-8502 \\
}

\recdate{December 22, 2003}

\abst{%
We investigate pairing symmetry and transition temperature 
in a quasi-one-dimensional repulsive Hubbard model. 
We solve the Eliashberg equation using the third-order perturbation 
expansion with respect to the on-site repulsion $U$. 
We find that when the electron number density is shifted from the 
half-filled state, 
a transition into unconventional superconductivity is expected. 
When one-dimensionality is weak, a spin-singlet state is favorable. 
In contrast, when one-dimensionality is strong and electron number density 
is far from the half-filled state, 
a spin-triplet state is stabilized. 
Finally, we discuss the possibility of unconventional superconductivity 
caused by the on-site Coulomb repulsion in $\beta$-Na$_{0.33}$V$_2$O$_5$. 
}

\kword{%
quasi-one-dimensional superconductors, pairing symmetry, 
transition temperature, third-order perturbation theory
}

\begin{document}
\sloppy
\maketitle

%\section{Introduction}
Superconductivity in quasi-one-dimensional conductors has been studied as an 
important phenomenon. 
Today, some quasi-one-dimensional superconductors, 
such as (TMTSF)$_2$X~\cite{rf:Ishiguro,rf:Jerome} 
and Sr$_{14-x}$Ca$_x$Cu$_{24}$O$_{41}$~\cite{rf:Uehara}, have been discovered, 
and their superconductivity has been investigated. 
Recently, a superconducting transition in $\beta$-Na$_{0.33}$V$_2$O$_5$, 
which has a quasi-one-dimensional lattice structure similar to 
Sr$_{14-x}$Ca$_x$Cu$_{24}$O$_{41}$, 
has been discovered~\cite{rf:Yamauchi}, and this phenomenon has attracted our 
attention. The transition temperature is $\Tc\simeq 8\,{\rm K}$ under high 
pressures of approximately $8\,$GPa. 
At ambient pressure, this material shows quasi-one-dimensional metallic 
behavior in an electric resistivity experiment at high 
temperature~\cite{rf:Yamada}, and encounters a charge-ordered transition 
at $T_{\rm CO}\simeq 135\,$K.~\cite{rf:Yamada}
Furthermore, below $T_{\rm N}\simeq 25\,{\rm K}$, a antiferromagnetic 
ordered phase appears in the charge-ordered phase.~\cite{rf:Ueda} 
Under high pressures of approximately 8$\,$GPa, the charge-ordered phase 
abruptly vanishes, accompanied by the superconducting 
transition~\cite{rf:Yamauchi}. 
It is not clear under high pressure whether 
the antiferromagnetic phase in the charge-ordered phase survives or not. 
However, the existence of the antiferromagnetic phase at ambient pressure 
suggests that the electron correlation is important. 

Such an electron correlation effect leads to unconventional superconductivity 
rather than conventional s-wave superconductivity induced by the 
electron-phonon coupling~\cite{rf:Yanase}. 
Such investigations have already been reported in the 
quasi-one-dimensional superconductors, (TMTSF)$_2$X and 
Sr$_{14-x}$Ca$_x$Cu$_{24}$O$_{41}$. 
The superconductivity in (TMTSF)$_2$X has been investigated using the 
fluctuation-exchange approximation (FLEX)~\cite{rf:Kino} and the third-order perturbation theory (TOPT)~\cite{rf:Nomura}. 
Both theoretical calculations suggest that a d-wave like spin-singlet state 
is the most stable. 
Also, in Sr$_{14-x}$Ca$_x$Cu$_{24}$O$_{41}$, the FLEX calculation for the 
trellis lattice indicated a d-wave like spin-singlet 
state~\cite{rf:Kontani}. 
However, experimentally, in both materials, the Knight shift does not change 
above and below $\Tc$, and the spin-triplet state is 
indicated~\cite{rf:Lee,rf:Fujiwara}, although it is still confusing. 
Thus, intensive investigations on quasi-one-dimensional superconductors 
should be carried out. 
In this letter, we investigate in detail unconventional superconductivity in 
a quasi-one-dimensional Hubbard model, taking up the superconductivity in 
$\beta$-Na$_{0.33}$V$_2$O$_5$. 
 
\begin{figure}[t]
\begin{center}
\includegraphics[width=8cm]{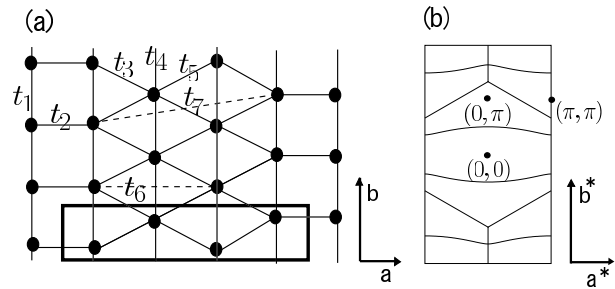}
\end{center}
\caption{(a) Schematic figure of the lattice used in this calculation. 
$t_i(1\leq i \leq 7)$ is the hopping integral. 
The region enclosed by rectangle is primitive 
cell. The primitive cell topologycally composes triangular lattice. 
(b) The Fermi surface for $n=0.90$, $t_0=1.0$. Since the lattice is 
topologycally triangular lattice, the Brillouin zone is hexagonal. }
\label{fig:lattice}
\end{figure}

%\section{Model}
Now, let us consider the lattice structure and the band structure in 
$\beta$-Na$_{0.33}$V$_2$O$_5$.
This material has three types of vanadium site, 
V$1$, V$2$ and V$3$. V$1$ is on the VO$_6$ zigzag chain, 
V$2$ is on the VO$_6$ ladder 
chain and V$3$ is on the VO$_5$ zigzag chain.
At ambient pressure, NMR experiments indicated 
that V3 does not have any conduction electrons
at low temperatures~\cite{rf:Itoh}. Assuming that conduction electrons at 
V3 are also empty under high pressures, we can consider the lattice 
structure with 
the V network shown in Fig. \ref{fig:lattice}(a), which is important for 
electric conductivity. This structure is different from the trellis lattice 
only in the V$1$ zigzag chain in the middle of Fig. \ref{fig:lattice}(a).
The unit cell is a thick-line rectangle, and contains four V sites. 
Since there is no information on the band structure of 
$\beta$-Na$_{0.33}$V$_2$O$_5$, we discuss a simple tight-binding model 
with an s orbital on each V site. 
In this case, we consider $7$ types of hopping integrals $(t_1\sim t_7)$ 
displayed in Fig. \ref{fig:lattice}(a).
By numerically diagonalizing the energy matrix, we obtain four bands, since 
there are four orbitals in the unit cell. 
Among them, we only use the lowest energy band. 
In $\beta$-Na$_{0.33}$V$_2$O$_5$, there is one electron per unit cell, if we 
simply count the valence electrons. 
Under the ideal condition that all electrons occupy only the lowest energy 
band, it becomes half-filled. 
Therefore, we deal with the electron number density $n$ as a parameter 
less than the half-filled state. 
If the superconductivity of this material is caused by electron correlation, 
electron correlation must be strong. 
For electron correlation to be strong, a high-density band is required. 
Therefore, the electrons should mainly occupy the lowest energy band. 
From these viewpoints, we use the single-band model. 
If there is no high-density band, we have to consider a mechanism other 
than electron correlation. 
As a typical set of parameters, we use $t_1=1.0$, $t_2=t_0$, 
$t_3=t_4=t_5=0.3t_0$ and $t_6=t_7=0.2t_0$. 
Here, $t_0$ is a measure of one-dimensionality, 
and with decreasing $t_0$, the Fermi surface becomes more one-dimensional. 
We assume that the results of calculations mainly depend on the 
one-dimensionality of the Fermi surface. 
Actually, when we change the ratio of the transfer integrals $t_i$ 
maintaing the one-dimensionality of the Fermi surface, 
the results of the calculations are almost unchanged. 
Therefore, we can study the dependence of the form of the Fermi 
surface using the parameter $t_0$. 
In Fig. \ref{fig:lattice}(b), we show a typical quasi-one-dimensional Fermi 
surface for $t_0=1.0$ and the electron number density $n=0.90$. 
This Fermi surface possesses a less nesting property, 
and is different from the band structures with an almost perfect 
nesting property discussed so far in the quasi-one-dimensional 
model calculations. 
Here, we investigate in detail superconductivity in such a situation. 
We consider the quasi-one-dimensional single-band Hubbard model with the 
lowest band $\varepsilon(k)$ discussed above, 
\begin{equation}
\begin{split}
H&=\sum_{k,\sigma}\varepsilon(k)c^{\dagger}_{k\sigma}
c_{k\sigma}\\
&+\frac{U}{2N}\sum_{k_i}\sum_{\sigma\neq\sigma'}
c^{\dagger}_{k_1\sigma}c^{\dagger}_{k_2\sigma'}c_{k_3\sigma'}
c_{k_4\sigma}\delta_{k_1+k_2,k_3+k_4}.
\end{split}
\end{equation}
We treat this model using the third-order perturbation expansion. 
Hereafter, in order to obtain a moderate transition temperature $\Tc$, 
we set $U=5.0$, which is almost equal to the bandwidth. 
The third-order perturbation theory in the strongly correlated region has 
been justified by higher-order calculations of pairing 
interactions.~\cite{rf:Nomura2} 
We can apply the perturbation theory for the appropriate values of $U$ to 
obtain the reliable value of $T_{\rm c}$. 
%\section{Formulation}

\begin{figure}[t]
\begin{center}
\includegraphics[width=8cm]{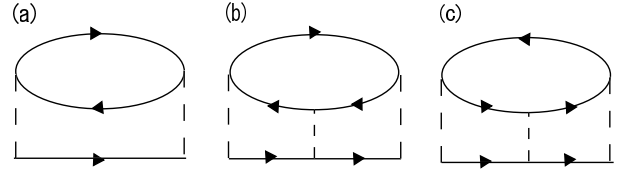}
\end{center}
\caption{The diagrams of Normal self-energy. The solid lines 
represent Green's function $G_0(k)$ and the broken lines represent 
the Coulomb repulsion $U$. 
}
\label{fig:selfenergy}
\end{figure}

We apply the third-order perturbation theory
with respect to $U$ to our model. 
Diagrams of the normal self-energy are 
shown in Fig. \ref{fig:selfenergy}. 
The normal self-energy is given by
\begin{equation}
\begin{split}
\Sigma_{\rm N}(k)&= \frac{T}{N}\sum_{k'} [U^2 \chi_0(k-k') G_0(k') \\
& +U^3 \left( \chi_0^2(k-k')+\phi_0^2(k+k') \right) G_0(k')], 
\end{split}
\end{equation}
where 
\begin{equation}
\begin{split}
&G_0(k)=\frac{1}{i\omega_n-\varepsilon (\ve{k})+\mu}, \\
&\chi_0(q)=-\frac{T}{N}\sum_{k} G_0(k)G_0(q+k), \\
&\phi_0(q)=-\frac{T}{N}\sum_{k}G_0(k)G_0(q-k).
\end{split}
\end{equation}
Here, $G_0(k)$ with the short notation 
$k=(\ve{k},\omega_n)$ represents the bare Green's function. 
Since the first-order normal self-energy is constant, it can be included by 
the chemical potential $\mu$.
The dressed Green's function $G(k)$ is given by
\begin{equation}
\begin{split}
G(k)=\frac{1}{i\omega_n-\varepsilon(\ve{k})-\Sigma_{\rm N}(k)+\mu+\delta\mu}. 
\end{split}
\end{equation}
Here, the chemical potential $\mu$ and the chemical potential shift 
$\delta\mu$ are determined so as to fix the electron number density $n$, 
\begin{equation}
\begin{split}
n=2\frac{T}{N}\sum_k G_0(k)=2\frac{T}{N}\sum_k G(k).
\end{split}
\end{equation}

We also expand the effective pairing interaction up to the third order 
with respect to $U$. 
For the spin-singlet state, the effective pairing interaction is given by
\begin{equation}
\begin{split}
V^{\rm Singlet}(k;k')=V_{\rm RPA}^{\rm Singlet}(k;k')
+V_{\rm Vertex}^{\rm Singlet}(k;k'),
\end{split}
\end{equation}
where
\begin{equation}
\begin{split}
V_{\rm RPA}^{\rm Singlet}(k;k')=U+U^2\chi_0(k-k')+2U^3\chi_0^2(k-k'), 
\end{split}
\end{equation}
and
\begin{equation}
\begin{split}
&V_{\rm Vertex}^{\rm Singlet}(k;k')=2(T/N)\Re\Big [\sum_{k_1}G_0(k_1) \\
&\times(\chi_0(k+k_1)-\phi_0(k+k_1))G_0(k+k_1-k')U^3\Big ].
\end{split}
\end{equation}

For the spin-triplet state,
\begin{equation}
\begin{split}
V^{\rm Triplet}(k;k')=V_{\rm RPA}^{\rm Triplet}(k;k')
+V_{\rm Vertex}^{\rm Triplet}(k;k'), 
\end{split}
\end{equation}
where
\begin{equation}
\begin{split}
V_{\rm RPA}^{\rm Triplet}(k;k')=-U^2\chi_0(k-k'), 
\end{split}
\end{equation}
and
\begin{equation}
\begin{split}
&V_{\rm Vertex}^{\rm Triplet}(k;k')=2(T/N)\Re\Big [\sum_{k_1}G_0(k_1)\\
&\times (\chi_0(k+k_1)+\phi_0(k+k_1))G_0(k+k_1-k')U^3\Big ].
\end{split}
\end{equation}
Here, $V_{\rm RPA}^{\rm Singlet(Triplet)}(k,k')$ is called the RPA terms and 
$V_{\rm Vertex}^{\rm Singlet(Triplet)}(k,k')$ is called the vertex 
corrections. 
Near the transition point, the anomalous self-energy $\Delta(k)$ satisfies 
the linearized Eliashberg equation,
\begin{equation}
\begin{split}
\lambda_{\rm max}\Delta(k)=-\frac{T}{N}\sum_{k'}V(k;k')|G(k')|^2\Delta(k'),
\end{split}
\end{equation}
where, $V(k;k)$ is $V^{\rm Singlet}(k;k')$ or $V^{\rm Triplet}(k;k')$, 
and $\lambda_{\rm max}$ is the largest positive eigenvalue. 
Then, the temperature at $\lambda_{\rm max}=1$ corresponds to $T_{\rm c}$.
By estimating $\lambda_{\rm max}$, 
we can determine which type of pairing symmetry is stable. 
For numerical calculations, we take 128 $\times$ 128 $\ve{k}$-meshes 
for twice space of the first Brillouin zone and 2048 Matsubara frequencies.

%\section{Obtain results}
%\subsection{Temperature dependence}

\begin{figure}[t]
\begin{center}
\includegraphics[width=8cm]{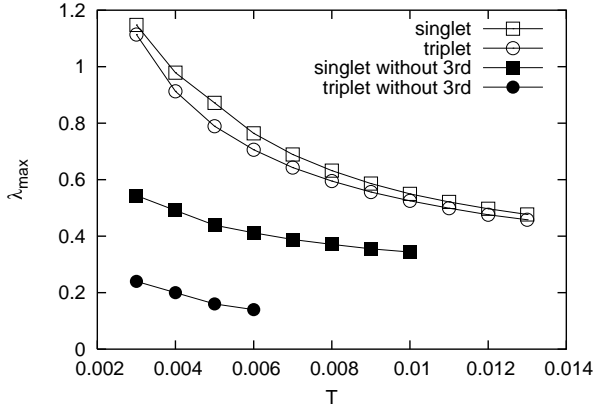}
\end{center}
\caption{Calculated maximum eigenvalues $\lambda_{\rm max}$ for 
spin-singlet 
(or spin-triplet) state. The line with white squares (circles) is the result 
for the spin-singlet (spin-triplet) state obtained using the third-order 
perturbation theory. The line with the black squares (circles) is the result 
for the spin-singlet 
(spin-triplet) state without the pairing interaction due to 
the third-order terms. The parameters are $n=0.90$ and $t_0=1.0$.}
\label{fig:temp}
\end{figure}

In Fig. \ref{fig:temp}, we show the results for 
$\lambda_{\rm max}$ 
in the case with $n=0.90$ and $t_0=1.0$. 
With decreasing temperature, $\lambda_{\rm max}$ increases. 
The spin-singlet state and the spin-triplet state possess almost the 
same transition temperature, $\Tc\simeq 0.004$.
If we assume that the bandwidth $W \simeq 5$ corresponds to $1\,{\rm eV}$, 
then $\Tc\simeq 8\,{\rm K}$ is obtained in accordance with 
the experimental value for $\beta$-Na$_{0.33}$V$_2$O$_5$.
In Fig. \ref{fig:temp}, we also show the results for $\lambda_{\rm max}$ 
obtained without the pairing interaction due to the third-order terms. 
For the spin-triplet state, we see that the vertex corrections are 
important for stabilizing the spin-triplet state from the comparison. 

%\subsection{Momentum dependence of $\chi_0(\ve{q},0),\phi_0(\ve{q},0)$}

\begin{figure}[t]
\begin{center}
\includegraphics[width=8cm]{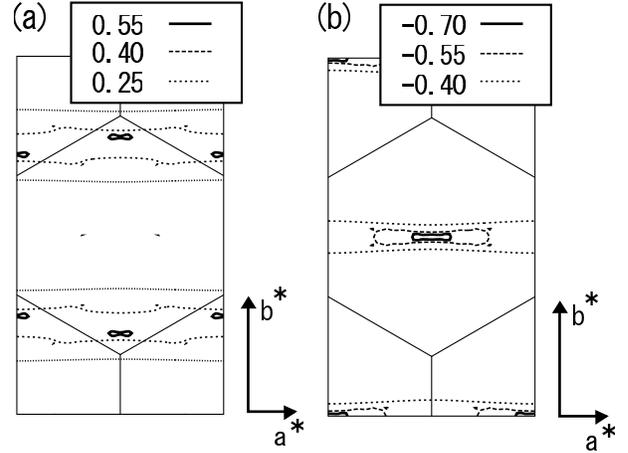}
\end{center}
\caption{(a) Contour plots of $\chi_0(\ve{q},0)$. The peaks exist near 
$\ve{q}=(\pi,\pi)$ 
and $\ve{q}=(0,\pi)$ 
(b) Contour plots of $\phi_0(\ve{q},0)$. The peaks exist at 
$\ve{q}=(0,0)$. The parameters are $n=0.90$, $t_0=1.0$ and $T=0.004$. }
\label{fig:chiphi}
\end{figure}

In Fig. \ref{fig:chiphi}, 
we show the momentum dependence of $\chi_0(\ve{q},0)$ and $\phi_0(\ve{q},0)$ 
in the case with $n=0.90$, $t_0=1.0$ and $T=0.004$. 
Since the Fermi surface has a nesting property, 
$\chi_0(\ve{q},0)$ has peaks near $\ve{q}=(\pi , \pi)$ and $\ve{q}=(0 , \pi)$. 
Since the Brillouin zone is hexagonal, $(\pi,\pi)$ is equivalent 
to $(0,\pi)$. 
At the half-filled state, $\chi_0(\ve{q},0)$ has peaks beside 
$\ve{q}=(\pi,\pi)$ and $\ve{q}=(0,\pi)$. 
If the electron number density $n$ is shifted from the half-filled state, 
then the peaks are shifted from $\ve{q}=(\pi,\pi)$ and $\ve{q}=(0,\pi)$ .
On the other hand, $\phi_0(\ve{q},0)$ has a peak at $\ve{q}=(0 , 0)$ . 
 
%\subsection{Momentum dependence of the anomalous self-energy}

\begin{figure}[t]
\begin{center}
\includegraphics[width=8cm]{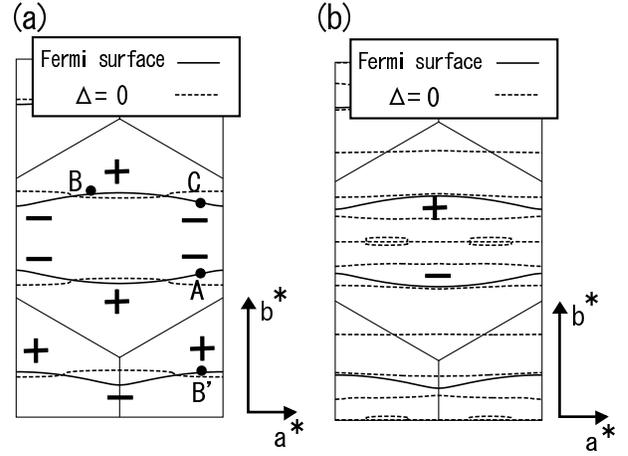}
\end{center}
\caption{(a) Contour plot of the anomalous self-energy. $\Delta(k)=0$ for the spin-singlet state has d-wave like momentum dependence. 
(b) Contour plot of the anomalous self-energy. $\Delta(k)=0$ for the 
spin-triplet state has p-wave like momentum dependence. The parameters are 
$n=0.90$, $t_0=1.0$ and $T=0.01$.}
\label{fig:gapsgapt}
\end{figure}

In Fig. \ref{fig:gapsgapt}, we show the contour plots of the anomalous 
self-energy in the case of $n=0.90$, $t_0=1.0$ and $T=0.01$. 
For the spin-singlet state, the momentum dependence of the anomalous 
self-energy on the Fermi surface is a d-wave like state with node. 
For the spin-triplet state, the momentum dependence of 
the anomalous self-energy is p-wave like, and is a fully gapped state. 
For the spin-singlet state, the RPA terms are dominant in the effective 
interaction terms. 
In this case, we can easily understand the gap structure in Fig. \ref{fig:gapsgapt}(a) from the structure in the Eliashberg equation as follows. 
The peak structures of $\chi_0(\ve{q},0)$ in Fig. \ref{fig:chiphi}(a) 
originate from the nesting property, between around point A and around 
point B, or around point B' on the Fermi surfaces 
in Fig. \ref{fig:gapsgapt}(a). 
In order to obtain a positive value of $\lambda_{\rm max}$, it is favorable 
that signs of $\Delta(k)$ around points B and B' are different from 
its sign around point A. 
The structure of $\Delta(k)$ in Fig. \ref{fig:gapsgapt}(a) just becomes so. 
%\subsection{Dependence of $\lambda_{\rm max}$ on the parameters $t_0$ and $n$}

\begin{figure}[t]
\begin{center}
\includegraphics[width=8cm]{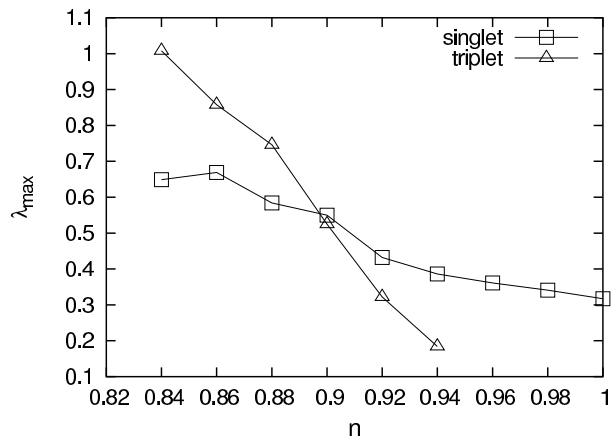}
\end{center}
\caption{$\lambda_{\rm max}$ 
as a function of $n$. The parameters are $T=0.01$ and $t_0=1.0$. } 
\label{fig:n}
\end{figure}

\begin{figure}[t]
\begin{center}
\includegraphics[width=8cm]{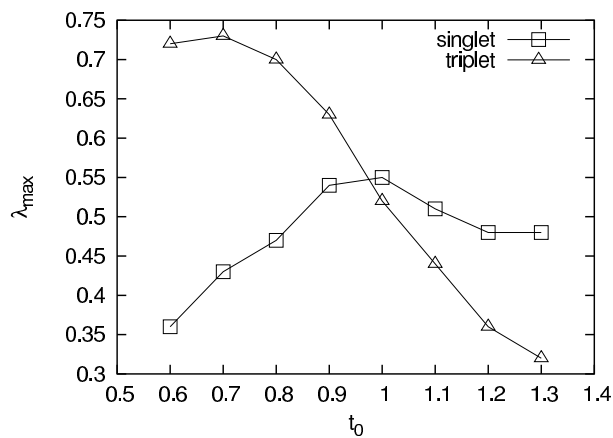}
\end{center}
\caption{$\lambda_{\rm max}$ as a function of $t_0$.
The parameters are $
n=0.90$ and $T=0.01$. }
\label{fig:t0}
\end{figure}
 
In Fig. \ref{fig:n}, 
$n$ dependence of $\lambda_{\rm max}$ is shown. 
If $n$ is near the half-filled state, the spin-singlet and the spin-triplet 
states have low values of $\lambda_{\rm max}$.
With decreasing $n$, $\lambda_{\rm max}$ increases, and 
the spin-singlet and spin-triplet states yield nearly the same 
$\lambda_{\rm max}$ 
at $n\simeq 0.90$. 
Moreover, with decreasing $n$ from $n=0.90$, the values of $\lambda_{\rm max}$ 
for the spin-triplet state become larger than 
those for the spin-singlet state. 
This indicates that the spin-triplet state may be realized far from the 
half-filled state. 
We can easily understand why the values of $\lambda_{\rm max}$ 
are suppressed for the spin-singlet and spin-triplet states at around 
the half-filled state. 
For spin-singlet state, at the half-filled state, 
$\chi_0(\ve{q},0)$ has a peak beside $\ve{q}=(0,\pi)$. 
Considering the structure of the Eliashberg equation, 
in order to obtain a large positive value of $\lambda_{\rm max}$, 
it is not favorable that signs of $\Delta(k)$ around point A are the same as 
its sign around point C on the Fermi surface in Fig. \ref{fig:gapsgapt}(a). 
Therefore, the values of $\lambda_{\rm max}$ are strongly suppressed around 
the half-filled state by the conflicting peaks of $\chi_0(\ve{q},0)$. 
On the other hand, if the electron number density 
is far from the half-filled state, the peak of 
$\chi_0(\ve{q},0)$ is far from $\ve{q}=(0,\pi)$. 
Therefore, the suppression of the values of $\lambda_{\rm max}$ becomes weak. 
For the spin-triplet state, 
when the Fermi surface has perfect particle-hole 
symmetry, the vertex corrections are
perfectly canceled out, and at approximately the half-filled state, owing to 
approximate particle-hole symmetry, vertex corrections are approximately 
canceled out and the values of $\lambda_{\rm max}$ are suppressed. 
Here, the normal self-energy term corresponding to 
Fig. \ref{fig:selfenergy}(c) make the mass enhancement factor small. 
When the Fermi surface have perfect particle-hole 
symmetry, the terms corresponding to Figs. \ref{fig:selfenergy}(b) 
and \ref{fig:selfenergy}(c) are 
perfectly canceled out. 
However, for $n\leq 0.82$, 
particle-hole symmetry deteriorates, 
and the mass enhancement factor is much smaller than unity. 
Therefore, reliable numerical calculation cannot be obtained in the range of 
$n\leq 0.82$.  

In Fig. \ref{fig:t0} we show the $t_0$ dependence of $\lambda_{\rm max}$. 
With decreasing $t_0$, the spin-triplet state becomes dominant, 
and with increasing $t_0$, the spin-singlet state becomes dominant. 
If $t_0$ is small, $\chi_0(\ve{q},0)$ have the character 
of one-dimensionality. 
In this case, the values of $\lambda_{\rm max}$  are suppressed 
by the conflict of the peaks of $\chi_0(\ve{q},0)$ like the above case. 

%\section{Discussions and Conclusion}
In conclusion, we have investigated pairing symmetry and the transition 
temperature on the basis of a quasi-one-dimensional Hubbard model. 
We have solved the Eliashberg equation using the third-order perturbation 
theory with respect to the on-site repulsion $U$. 
We find that if $n$ is shifted from the 
half-filled state, 
the transitions into unconventional superconductivity is expected. 
If one-dimensionality is weak, a spin-singlet pairing is more stable 
than a spin-triplet one. 
In contrast, if one-dimensionality is strong and $n$ is far from the half-filled state, a spin-triplet pairing is more stable than 
a spin-singlet one. 
Thus, we suggest the possibility of unconventional superconductivity in 
$\beta$-Na$_{0.33}$V$_2$O$_5$ 
caused by the on-site Coulomb repulsion.
%\section{Acknowledgements}

Numerical calculation in this work was carried out at 
the Yukawa Institute Computer Facility.
%%%%%%%%%%%%%%%%%%References%%%%%%%%%%%%%%%%%%%

\end{document}